\documentclass[10pt,onecolumn]{article}
\usepackage{times,bm}
\usepackage[affil-it,]{authblk}
\usepackage[pagebackref=false,colorlinks=true,linkcolor=red,citecolor=blue,urlcolor=black]{hyperref}
\usepackage{geometry}
\usepackage[switch*,pagewise, mathlines]{lineno}
\usepackage{lineno}
\usepackage{graphicx}
\usepackage{cite}
\usepackage{amsmath}
\usepackage{amsfonts}
\usepackage{amssymb}
\usepackage{slashed}
\usepackage{color}
\usepackage{multirow,multicol}
\usepackage{tabulary}
\usepackage[flushleft]{threeparttable}
\usepackage{pgfplots}
\pgfplotsset{compat=1.18}
\usepackage{xcolor}
\numberwithin{equation}{section}
\usepackage{tikz}
\usepackage{ulem}
\usepackage{hyperref}
\usepackage{nameref}
\usepackage{comment}

\usepgflibrary{arrows}
\usetikzlibrary{shapes.callouts}
\tikzset{
	level/.style   = { thick, },
	connect/.style = { dotted, red   },
	notice/.style  = { draw, rectangle callout, callout relative pointer={#1} },
	label/.style   = { text width=2cm }
}

\definecolor{acsblue}{RGB}{39,117,215}
\definecolor{shadecolor}{RGB}{255,241,204}

\usepackage{letltxmacro}
\makeatletter
\let\oldr@@t\r@@t
\def\r@@t#1#2{%
	\setbox0=\hbox{$\oldr@@t#1{#2\,}$}\dimen0=\ht0
	\advance\dimen0-0.2\ht0
	\setbox2=\hbox{\vrule height\ht0 depth -\dimen0}%
	{\box0\lower0.4pt\box2}}
\LetLtxMacro{\oldsqrt}{\sqrt}
\renewcommand*{\sqrt}[2][\ ]{\oldsqrt[#1]{#2}}
\makeatother
\usepackage{environ}
\begin{document}

\newcommand{{\ri}}{{\rm{i}}}
\newcommand{{\Psibar}}{{\bar{\Psi}}}

\title{Vector bosons in the rotating frame of \textcolor{black}{negative curvature} wormholes}
\author{ \textit {Abdullah Guvendi}$^{\ 1}$\footnote{\textit{E-mail: abdullah.guvendi@erzurum.edu.tr } }~,~ \textit {Semra Gurtas Dogan}$^{\ 2}$\footnote{\textit{E-mail: semragurtasdogan@hakkari.edu.tr (Corr. Auth.)} }  \\
	\small \textit {$^{\ 1}$ Department of Basic Sciences, Faculty of Science, Erzurum Technical University, 25050, Erzurum, Turkey}\\
	\small \textit {$^{\ 2}$ Department of Medical Imaging Techniques, Hakkari University, 30000, Hakkari, Turkey}\\}

\date{}
\maketitle

\begin{abstract}
\textcolor{black}{In this study, we investigate the relativistic dynamics of vector bosons within the context of rotating frames of negative curvature wormholes. We seek exact solutions for the fully-covariant vector boson equation, derived as an excited state of zitterbewegung. This equation encompasses a symmetric rank-two spinor, enabling the derivation of a non-perturbative second-order wave equation for the system under consideration. Our findings present exact results in two distinct scenarios. Notably, we demonstrate the adaptability of our results to massless vector bosons without compromising generality. The evolution of this system is shown to correlate with the angular frequency of the uniformly rotating reference frame and the curvature radius of the wormholes. Moreover, our results highlight that the interplay between the spin of the vector boson and the angular frequency of the rotating frame can give rise to real oscillation modes, particularly evident in excited states for massless vector bosons. Intriguingly, we note that the energy spectra obtained remain the same whether the wormhole is of hyperbolic or elliptic nature.}
  \end{abstract}

\begin{small}
\begin{center}
\end{center}
\end{small}

\section{Introduction}\label{sec1}

\textcolor{black}{The pursuit of solving fully-covariant wave equations for relativistic quantum particles in curved spaces holds immense significance in theoretical physics. This quest unites quantum mechanics with general relativity, offering a fundamental framework to comprehend particle and field behavior in such environments \cite{parker}. It may represent a pivotal step towards formulating a theory of quantum gravity, shedding light on the quantum essence of gravity itself. Accurate solutions to these equations bear potential for experimental verification, offering insights into the interplay between quantum mechanics and gravitational fields \cite{qg-1,qg-2,qg-3,qg-4,qg-5}. Moreover, investigating vector bosons, carriers of fundamental forces, in curved spaces is crucial because understanding how these bosons interact with such environments may aid in comprehending the impact of gravitational fields on fundamental forces. Additionally, exploring the effects of rotating reference frames on quantum systems provides profound insights into the interconnections between quantum mechanics and relativity \cite{ref2,ref3,ref4,ref5}. This investigation elucidates the transformation of quantum states, the complexities introduced by non-commutative quantum operators, and the influences of non-inertial effects on observed energy levels and phase shifts \cite{exp1,exp2,exp3,ref6,ref7,ref8,ref9,ref10,ref11,ref12,ref13,ref14,ref15,ref16,ref17,ref18,ref19,ref20}. Such studies, when tested through experiments, hold the key to validating theoretical predictions, and deepening our understanding of quantum mechanics, relativity, and their potential technological applications \cite{exp1,exp2,exp3}. This paper aims to derive analytical results for relativistic spin-1 particles in the rotating frame of \textcolor{black}{negative curvature} wormholes \cite{wormhole,gibbons} through the fully-covariant vector boson equation established by Barut \cite{barut}.}

\textcolor{black}{Barut elucidated a unifying principle that systematically derives the widely recognized fully-covariant wave equations governing the dynamics of spinning particles} \cite{barut}. The \textcolor{black}{vector boson equation} was introduced as an excited state of zitterbewegung, and it corresponds to the spin-1 sector of the Duffin-Kemmer-Petiau equation in (2+1)-dimensions \cite{cavit,ganim,guvendi2022vector,vbo1,vbo2,vbo3}. The corresponding spinor is constructed through the direct product of \textcolor{black}{symmetric two Dirac spinors}, so the \textcolor{black}{vector boson equation} includes a symmetric spinor of rank-two \cite{barut}. \textcolor{black}{This facilitates the derivation of non-perturbative outcomes applicable across various physical systems. It proves beneficial to highlight various types of research in this context.} Introducing a quantum analogy of Schumann resonances \cite{cavit}, investigating quantum tunnelling properties of a massive spin-1 particles from the Warped-Ad$S_{3}$ black holes \cite{ganim}, determining the evolution of relativistic spin-1 oscillator field in the near-horizon region of black holes \cite{guvendi2022vector}, analyzing the effects of stable one-dimensional topological defects on the generalized  \textcolor{black}{vector boson} oscillator \cite{vbo1,vbo2,vbo3} can be considered among them. However, we could not find any announced result for \textcolor{black}{vector bosons} in \textcolor{black}{wormholes} or in the \textcolor{black}{rotating frame} of \textcolor{black}{wormholes}. To fill this gap and discuss many interesting effects, we will try to explore the evolution of relativistic vector bosons in the \textcolor{black}{rotating frame} of the \textcolor{black}{negative curvature wormholes} in two different scenarios by solving the corresponding form of the \textcolor{black}{vector boson equation}.

\textcolor{black}{This manuscript is organized as follows: In Section \ref{sec2}, we commence by presenting the generalized form of the \textcolor{black}{vector boson equation} and proceed by deriving a system of equations comprising three equations. Section \ref{sec3} is dedicated to unveiling a comprehensive non-perturbative second-order wave equation applicable to the systems under consideration. Consequently, we attain precise solutions for relativistic \textcolor{black}{vector bosons} in the \textcolor{black}{rotating frame} of both the hyperbolic wormhole and elliptic wormhole in Sections \ref{sec3.1} and \ref{sec3.2}, respectively. Section \ref{sec4} encapsulates a concise summary of our findings followed by a discussion of the obtained results across several physically plausible scenarios.}

\section{\textcolor{black}{Generalized vector boson equation in the rotating frame of negative curvature wormholes}} \label{sec2}

In this part, we introduce the generalized form of the \textcolor{black}{vector boson equation} in a (2+1)-dimensional curved space, and then we derive a $3\times 3$ dimensional matrix equation for relativistic \textcolor{black}{vector bosons} in the \textcolor{black}{rotating frame} of \textcolor{black}{negative curvature wormholes described by} 2-dimensional curved surface of constant negative Gaussian curvature. In (2+1)-dimensional curved spacetime, the \textcolor{black}{vector boson equation} can be written as the following \cite{guvendi2022vector}
\begin{eqnarray}
\left\lbrace \mathcal{B}^{\mu}\slashed{\nabla}_{\mu}+i\tilde{m}\textbf{I}_{4}\right\rbrace \Psi\left(\textbf{x}\right)=0,   \label{eq1}
\end{eqnarray}
in which the Greek indices refer to curved spacetime coordinates, $\slashed{\nabla}_{\mu}$ denote covariant derivatives, $\slashed{\nabla}_{\mu}=\partial_{\mu}-\Omega_{\mu}$, \textcolor{black}{$\mathcal{B}^{\mu}$} are the space-dependent matrices constucted through the generalized Dirac matrices ($\gamma^{\mu}$) in such a way that $\mathcal{B}^{\mu}=\frac{1}{2}\left[\gamma^{\mu}\otimes \textbf{I}_{2}+\textbf{I}_{2}\otimes\gamma^{\mu}\right]$, $\tilde{m}=\frac{mc}{\hbar}$, and $\Psi\left(\textbf{x}\right)$ is the spacetime-dependent symmetric spinor. Here, $\textbf{I}_{2}$($\textbf{I}_{4}$) stand for the 2(4)-dimensional identity matrices, $m$ is the rest mass of the \textcolor{black}{vector boson}, $c$ is the light speed, $\hbar$ is the reduced Planck constant, the symbols $\otimes$ indicate the Kronecker product, $\textbf{x}$ is the spacetime position vector, and $\Omega_{\mu}$ are the spinorial affine connections that can be constructed in terms of the the \textcolor{black}{spinorial affine connections} ($\Gamma_{\mu}$) for Dirac field, as $\Omega_{\mu}=\Gamma_{\mu}\otimes \textbf{I}_{2}+\textbf{I}_{2}\otimes \Gamma_{\mu}$. The generalized Dirac matrices are obtained through $\gamma^{\mu}=e^{\mu}_{k}\gamma^{k}$ where the Latin index $k$ refers to coordinates of the flat Minkowski spacetime ($k=0,1,2.$),  $e^{\mu}_{k}$ are the inverse tetrad fields, and $\gamma^{k}$ are the space-independent Dirac matrices chosen in terms of the Pauli spin matrices ($\sigma^{x}, \sigma^{y}, \sigma^{z}$) in (2+1)-dimensions. The inverse tetrads can be determined by using the following relation: $e^{\mu}_{k}=g^{\mu\nu}e_{\nu}^{l}\eta_{k l}$ in which $g^{\mu\nu}$ is the contravariant metric tensor, \textcolor{black}{$e_{\nu}^{l}$} are the tetrad fields, and $\eta_{k l}$ stands for the flat Minkowski tensor with the signature $(+,-,-)$, that is $\eta_{k l}=\textrm{diag}(1,-1,-1)$. For this choice of the signature, the flat Dirac matrices can be chosen as $\gamma^{0}=\sigma^{z}$, $\gamma^{1}=i\sigma^{x}$ and $\gamma^{2}=i\sigma^{y}$, in which $i=\sqrt{-1}$, since $\sigma_{x(yz)}^{2}=\textbf{I}_{2}$. The tetrad fields $e_{\nu}^{l}$ are determined by using the covariant metric tensor ($g_{\mu\nu}$) as the following $g_{\mu\nu}=e_{\mu}^{k}e_{\nu}^{l}\eta_{k l}$. Also, the spinorial connections for Dirac field can be determined through $\Gamma_{\lambda}=\frac{1}{4}g_{\mu\epsilon}\left[e^{k}_{\nu,\lambda}e^{\epsilon}_{k}-\Gamma_{\nu \lambda}^{\epsilon}\right]\mathcal{S}^{\mu\nu}$ in which $_{,\lambda}$ denotes derivative with respect to coordinate \textcolor{black}{$x^{\lambda}$}, $\Gamma_{\nu\lambda}^{\epsilon}$ are the Christoffel symbols, \textcolor{black}{$\Gamma_{\nu\lambda}^{\epsilon}=\frac{1}{2}g^{\epsilon \alpha}\left[\partial_{\nu} g_{\lambda \alpha}+\partial_{\lambda} g_{\alpha \nu}-\partial_{\alpha} g_{\nu\lambda} \right]$}, and $\mathcal{S}^{\mu\nu}$ are the spin operators, namely $\mathcal{S}^{\mu\nu}=\left[\gamma^{\mu},\gamma^{\nu} \right]$.
Now, let us introduce a spacetime background describing the \textcolor{black}{negative curvature wormholes} known also as \textcolor{black}{hyperbolic wormhole} and \textcolor{black}{elliptic wormhole \cite{wormhole}}. Such a spacetime background can be represented through the following metric \cite{wormhole,gibbons}
\begin{eqnarray*}
ds^{2}=c^2dT^2-du^2-\chi^{2}\left(u\right)dv^2
\end{eqnarray*}
where $\chi\left(u\right)$ is
\begin{itemize}
\item $\chi\left(u\right)= a\ \textrm{cosh}\left(u/\rho_{0}\right)$ for \textcolor{black}{hyperbolic wormhole},
\item $\chi\left(u\right)= b\ \textrm{sinh}\left(u/\slashed{\rho}_{0}\right)$ for \textcolor{black}{elliptic wormhole}.
\end{itemize}
Here, $\textcolor{black}{a\ (b)}$  stands for the radius of the \textcolor{black}{wormhole} at the mid-point ($u=0$) between two ends, $\textcolor{black}{\rho_{0}\  (\slashed{\rho}_{0})}$  is the radius of the curvature of the \textcolor{black}{wormhole} surface along \textcolor{black}{$u$}. Also, it is worth underlining that the Gaussian curvature ($\mathcal{K}$) for the considered \textcolor{black}{wormhole} backgrounds is $\mathcal{K} =-\chi_{,uu}/\chi$ \cite{wormhole,gibbons}. Now, we can introduce a general line element describing the rotating reference frame of these \textcolor{black}{negative curvature wormholes}. This can be acquired through the following coordinate transformations: $T\longrightarrow t$, $u\longrightarrow \rho$ and \textcolor{black}{$v\longrightarrow \phi+ \omega_{rf} T$ where $\omega_{rf}$ is the angular frequency of the uniformly rotating frame}. At that rate, the metric tensor describing the uniformly \textcolor{black}{rotating frame of negative curvature wormholes} can be determined as
\textcolor{black}{\begin{flalign*}
g_{\mu\nu}=\left(\begin{array}{ccc}
c^2-\omega_{rf}^{2} \chi^2& 0 & -\omega_{rf} \chi^2\\
0& -1 & 0\\
-\omega_{rf} \chi^2& 0 & - \chi^2
\end{array}\right),
\end{flalign*}}
for which one can obtain the contravariant metric tensor as the following
\textcolor{black}{\begin{flalign*}
g^{\mu\nu}=\left(\begin{array}{ccc}
1/c^2& 0 & -\omega_{rf}  /c^2 \\
0& -1 & 0\\
-\omega_{rf}  /c^2 & 0 & \omega_{rf}^{2}  /c^2- 1/ \chi^2
\end{array}\right),
\end{flalign*}}
since $g_{\mu\nu}g^{\mu\nu}=\textbf{I}_{3}$ where \textcolor{black}{$x^{\lambda}=t,\rho,\phi$}. According to the metric tensors, we can obtain the non-vanishing components of the Christoffel symbols as the following
\textcolor{black}{\begin{eqnarray*}
&\Gamma_{tt}^{\rho}=-\omega_{rf}^{2}\chi\chi_{,\rho},\quad \Gamma_{t\phi}^{\rho}=-\omega_{rf}\chi\chi_{,\rho},\quad \Gamma_{\phi\phi}^{\rho}=-\chi\chi_{,\rho}\nonumber\\
&\Gamma_{t\rho}^{\phi}=\omega_{rf}\chi_{,\rho}/\chi ,\quad \Gamma_{\rho\phi}^{\phi}=\chi_{,\rho}/\chi.
\end{eqnarray*}}
Accordingly, one obtains the generalized Dirac matrices ($\gamma^{\mu}$), and \textcolor{black}{non-vanishing} spinorial affine connections, $\Gamma_{\mu}$, as the following
\textcolor{black}{\begin{eqnarray}
&\gamma^{t}=\frac{1}{c} \sigma^{z},\quad \gamma^{\rho}=i \sigma^{x},\quad \gamma^{\phi}=-\frac{\omega_{rf}}{c}\, \sigma^{z}+\frac{i}{\chi} \sigma^{y},\nonumber\\
&\Gamma_{t}= \frac{i}{2} \omega_{rf}\, \chi_{,\rho} \sigma^{z},\quad \Gamma_{\phi}= \frac{i}{2} \chi_{,\rho} \sigma^{z},   \label{eq2}
\end{eqnarray}}
since
\textcolor{black}{\begin{eqnarray*}
e^{\mu}_{k}=\left(\begin{array}{ccc}
1/c& 0 & 0 \\
0& 1 & 0\\
-\omega_{rf}/c & 0 & 1/ \chi
\end{array}\right).
\end{eqnarray*}}
By factorising the \textcolor{black}{symmetric} spinor $\Psi\left(\textbf{x}\right)=\textcolor{black}{\textrm{e}^{-i\frac{\mathcal{E}}{\hbar} t}}\textrm{e}^{is \phi}\left(\psi_{1} \ \psi_{2}\  \psi_{3}\  \psi_{4}\right)^{\textbf{T}}$, \textcolor{black}{in which $\mathcal{E}$ is the relativistic energy}, $s$ is the spin, $\textbf{T}$ means transpose of the $\rho$-dependent part of the spinor and substituting the results (\ref{eq2}) into the Eq. (\ref{eq1}), and then performing some arrangements one can derive the following $3\times 3$-dimensional matrix equation
\textcolor{black}{\begin{eqnarray*}
\left(\begin{array}{ccc}
\tilde{\lambda}& -s/\chi & -\tilde{m} \\
-\tilde{m}& -\partial_{\rho} & \tilde{\lambda}\\
\partial_{\rho}+\chi_{,\rho}/\chi & \tilde{m} & -s/\chi
\end{array}\right)\left(\begin{array}{ccc}
\psi_{+}\\
\psi_{0}\\
\psi_{-}
\end{array}\right)=0.
\end{eqnarray*}}
Here, $\tilde{\lambda}=\textcolor{black}{\frac{\mathcal{E}}{\hbar\,c}}+s\,\omega_{rf}/c$, $\psi_{\pm}=\psi_{1}\pm \psi_{4}$ and $\psi_{0}=2 \psi_{2}$ due to $\psi_{2}=\psi_{3}$ (see also \cite{guvendi2022vector}). \textcolor{black}{The matrix equation results in a set of three equations, one of which is algebraic\footnote{\textcolor{black}{In this set of equations, the first equation is algebraic.}}
\begin{eqnarray}
&\tilde{\lambda}\psi_{+}-\frac{s}{\chi}\psi_{0}-\tilde{m}\psi_{-}=0,\nonumber\\
&\tilde{\lambda}\psi_{-}-\tilde{m}\psi_{+}-\partial_{\rho}\psi_{0}=0,\nonumber\\
&\tilde{m}\psi_{0}+\left[\partial_{\rho}+\frac{\chi_{,\rho}}{\chi}\right]\psi_{+}-\frac{s}{\chi}\psi_{-}=0.\label{eq3}
\end{eqnarray}
These equations allow us to write the defined components $\psi_{+}$ and $\psi_{-}$ in terms of the $\psi_{0}$ as the following
\begin{flalign}
\psi_{-}=\frac{\tilde{\lambda}\tilde{m}}{\tilde{\lambda}^2-\tilde{m}^2}\left[\frac{1}{\tilde{m}}\psi_{0_{,\rho}}+\frac{s}{\tilde{\lambda}\chi}\psi_{0}\right], \quad \psi_{+}=\frac{\tilde{m}^2}{\tilde{\lambda}^2-\tilde{m}^2}\left[\frac{1}{\tilde{m}}\psi_{0_{,\rho}}+\frac{s}{\tilde{\lambda}\chi}\psi_{0}\right]+\frac{s}{\tilde{\lambda}\chi}\psi_{0}.\label{components}
\end{flalign}
Here, it is worth underlining that the Eq. (\ref{components}) allows us to recover the components $\psi_{1}$ and $\psi_{4}$ in terms of the $\psi_{2}$, and accordingly we can obtain the following expressions\footnote{\textcolor{black}{By using these expressions, we will write the spinor $\Psi\left(\textbf{x}\right)$ in explicit form.}}
\begin{eqnarray}
&\psi_{1}=\frac{1}{\tilde{\lambda}-\tilde{m}}\left[\psi_{2_{,\rho}}+\frac{s}{\chi}\psi_{2}\right],\quad \psi_{4}=-\frac{1}{\tilde{\lambda}+\tilde{m}}\left[\psi_{2_{,\rho}}-\frac{s}{\chi}\psi_{2}\right].\label{f-comp}
\end{eqnarray}
By substituting the results in the Eq. (\ref{components}) into the third equation in the Eq. (\ref{eq3}) leads to a general non-perturbative second-order wave equation in terms of the $\psi_{0}\left(\rho\right)$. In the following section, we will derive the explicit form of this wave equation and explore exact solutions in two distinct scenarios.}

\section{Exact results}\label{sec3}

\textcolor{black}{In this section, let us start by writing explicit the form of the aforementioned non-perturbative wave equation. By solving the equations in the Eq. (\ref{eq3}) for the component $\psi_{0}$, one obtains the following wave equation
\begin{eqnarray}
\left[\partial_{\rho}^2+\frac{\chi_{,\rho}}{\chi}\partial_{\rho}+\varepsilon-\frac{s^2}{\chi^2}\right]\psi_{0}\left(\rho \right) =0,\label{eq4}
\end{eqnarray}
where $\varepsilon =\tilde{\lambda}^2-\tilde{m}^2$.}

\subsection{\textcolor{black}{Vector bosons in the rotating frame of hyperbolic wormhole}}\label{sec3.1}

Here, first of all, we consider the relativistic \textcolor{black}{vector boson} in the \textcolor{black}{rotating frame} of the \textcolor{black}{hyperbolic wormhole}. In this scenario, the Eq. (\ref{eq4}) becomes as the following
\begin{eqnarray}
\left[\partial_{\rho}^2+\frac{\textrm{tanh}\left(\frac{\rho}{\rho_{0}}\right)}{\rho_{0}}\partial_{\rho}+\varepsilon-\frac{s^2}{a^2 \textrm{cosh}^{2}\left(\frac{\rho}{\rho_{0}} \right) }\right]\psi_{0}=0.\label{eq5}
\end{eqnarray}
This seemingly unfamiliar wave equation can be reduced to a familiar one by considering a new change of variable, $z=\sqrt{1-\textrm{cosh}^{2}\left(\rho/\rho_{0} \right)}$ \footnote{Note that $z\longrightarrow 0$ when $\rho\longrightarrow0$.}. In terms of the variable $z$, the Eq. (\ref{eq5}) can be written as in the following form
\begin{eqnarray}
\left(1-z^2 \right)\psi_{0_{,zz}}-2z\psi_{0_{,z}}+\left[\textcolor{black}{\xi\left(\xi+1 \right)}-\frac{ \varphi^{2} }{1-z^2} \right]\psi_{0}=0,  \label{eq6}
\end{eqnarray}
where
\begin{eqnarray*}
\xi=\frac{1}{2}\sqrt{1-4\varepsilon\rho^{2}_{0}}-\frac{1}{2},\quad \varphi=\frac{is\rho_{0}}{a}.
\end{eqnarray*}
The Eq. (\ref{eq6}) is the well-known associated Legendre differential equation and its regular solution around the origin can be expressed in terms of the associated Legendre function ($\mathcal{P}^{\varphi}_{\xi}$), as $\psi_{0}\left(z\right)=\mathcal{C}\,\mathcal{P}^{\varphi}_{\xi}\left(z \right)$ where $\mathcal{C}$ is an arbitrary constant. The function $\mathcal{P}^{\varphi}_{\xi}\left(z\right)$ becomes polynomial of degree $n$ with respect to $z$ if and only if $\xi=n$ where $n$ is the radial quantum number ($n=0,1,2...$). This condition results in the quantization condition for the energy \textcolor{black}{($\mathcal{E}\longrightarrow \mathcal{E}_{ns}$)} of the system in question. Accordingly, one can obtain the following energy spectra

\textcolor{black}{
\begin{eqnarray}
\mathcal{E}_{ns}=-s \hbar\omega_{rf}  \pm \hbar c \sqrt{\frac{m^2c^2}{\hbar ^{2}}-\frac{n\left( n+1\right) }{\rho_{0}^2}}.    \label{eq7}
\end{eqnarray}}

\textcolor{black}{Now, according to findings, we can easily write the symmetric spinor describing the considered vector field (spin-1 field) in the Eq. (\ref{eq1}) in terms of the argument $z$
\begin{eqnarray}
\Psi=\textrm{e}^{-i\frac{\mathcal{E}_{ns}}{\hbar} t}\textrm{e}^{is\phi}\,\mathcal{C}\,\left(\begin{array}{c}
-\frac{\sqrt{z^2-1}}{\rho_{0}(\tilde{\lambda}-\tilde{m})}\left[\partial_{z}\,\mathcal{P}^{\varphi}_{\xi}(z)\right]+\frac{is/a}{\sqrt{z^2-1}}\mathcal{P}^{\varphi}_{\xi}(z) \\
\mathcal{P}^{\varphi}_{\xi}(z)\\
\mathcal{P}^{\varphi}_{\xi}(z) \\
\frac{\sqrt{z^2-1}}{\rho_{0}(\tilde{\lambda}+\tilde{m})}\left[\partial_{z}\,\mathcal{P}^{\varphi}_{\xi}(z)\right]-\frac{is/a}{\sqrt{z^2-1}}\mathcal{P}^{\varphi}_{\xi}(z)
\end{array}\right).\label{field-HWH}
\end{eqnarray}}
\textcolor{black}{Here, it is clear that the energy of such a system does not depend on the radius of the wormhole. However, the solution function depends explicitly on the radius of the wormhole. Let's examine Eq. (\ref{eq6}) now. If we consider massless vector bosons, our results may be useful still because shape of this equation does not change when $m^{2}\rightsquigarrow 0$. In this limit, the system's energy takes on the following characteristics \textcolor{black}{$\mathcal{E}_{ns}\rightsquigarrow-s \hbar\omega_{rf}  \pm i\,\hbar c \sqrt{\frac{n\left( n+1\right) }{\rho_{0}^2}}$}. For the ground state ($n=0$) of such a system, it is very interesting that one may measure only the energy contribution $(\propto s \hbar\omega_{rf})$ stemming from the coupling of the particle's spin with the angular frequency of the uniformly \textcolor{black}{rotating frame} and moreover this energy value changes according to the angular frequency of the \textcolor{black}{rotating frame} as well as the physically possible spin quantum states $s=0, \pm 1$. Here, it is also clear that the term $s \hbar\omega_{rf}$ can be responsible for symmetry breaking around the zero energy if $m^{2}> 0$. Also, one can realize that the energy of such system becomes $\mathcal{E}_{00}\rightsquigarrow 0$ if $m^{2}\rightsquigarrow 0$. If we consider massless vector bosons, it is clear that the system may possess damped modes besides the real oscillations especially for the excited states. If such a quantum state can occur, the system cannot be stable and the corresponding modes decay (or grow) exponentially in time (note that $\Psi\propto\textrm{e}^{-i\frac{\mathcal{E}_{ns}}{\hbar} t}$) with a lifetime $\tau_{n}=\frac{\hbar}{|\mathcal{E}_{n_{Im}}|}$\footnote{Here, $\mathcal{E}_{n_{Im}}$ is the imaginary part of the energy.} \cite{guvendi2022vector}
\begin{eqnarray*}
\tau_{n}\rightsquigarrow \frac{\rho_{0}}{c\sqrt{n(n+1)}},
\end{eqnarray*}
provided $n\geqslant 1$ and $m^{2}\rightsquigarrow 0$. Here, one should note that the $\rho_{0}$ is in units of length. In principle, it seems possible to tune the evolution of vector bosons in the rotating frame of hyperbolic wormhole. Furthermore, it is essential to emphasize that the energy of a vector boson (spin-1 field) remains unaffected by the radius $a$ of the wormhole; however, it is evident that the radius $a$ influences the wave function's behavior. In principle, it is possible to manage the time evolution of relativistic spin-1 fields within the rotating frame of the hyperbolic wormhole by adjusting the radius of curvature ($\rho_{0}$) of the wormhole since such structures can be rolled, twisted, and curved \cite{wormhole}. In this scenario, we can also conclude that the lifetime of a vector boson carrying mass $m \lesssim \frac{\hbar}{\rho_{0}c}\sqrt{n(n+1)}$ in the hyperbolic wormhole can be extremely long for the excited states. However, we know that the considered system will reach to ground state eventually. The energy of the ground state is contingent upon the angular frequency of the rotating frame and the spin polarization $(s=0, \pm 1)$ of the vector field, devoid of any reflection of curvature effects, regardless of whether $m = 0$ or not.
Here, it is worth underlining that the spin-1 field (vector boson) has three polarization states, which can be categorized into two types: longitudinal and transverse components. The longitudinal component corresponds to the polarization of the spin-1 particle along its direction of motion. In terms of the vector field, the longitudinal polarization ($s=0$) corresponds to the component parallel to the particle's momentum. The transverse components ($s=\pm 1$) of the spin-1 particle's polarization are perpendicular to its direction of motion. In terms of the vector field, these transverse polarizations correspond to the components perpendicular to the particle's momentum. For massless vector bosons (photons), transverse polarizations are associated with the two physical polarizations of light that can be observed experimentally.}

Now, let us consider a curved surface describing the \textcolor{black}{rotating frame} of \textcolor{black}{elliptic wormhole}.

\subsection{\textcolor{black}{Vector bosons in the rotating frame of elliptic wormhole}}\label{sec3.2}

In this part, we are interested in non-inertial effects on the relativistic \textcolor{black}{vector bosons} in \textcolor{black}{elliptic wormhole}. In this case, the Eq. (\ref{eq4}) can be written as follows
\begin{eqnarray}
\left[\partial_{\rho}^2+\frac{\textrm{coth}\left(\frac{\rho}{\slashed{\rho}_{0}}\right)}{\slashed{\rho}_{0}}\partial_{\rho}+\varepsilon-\frac{s^2}{b^2 \textrm{sinh}^{2}\left(\frac{\rho}{\slashed{\rho}_{0}} \right) }\right]\psi_{0}=0.\label{eq8}
\end{eqnarray}
Here, it is clear that one needs to get rid of the hyperbolic functions to derive a familiar, and soluble wave equation. This may be acquired through a new change of variable, reads as $x=\textrm{cosh}\left(\rho/\slashed{\rho}_{0}\right)$. In terms of $x$, the Eq. (\ref{eq5}) can be written as in the following form
\begin{eqnarray}
\left(1-x^2 \right)\psi_{0_{,xx}}-2x\psi_{0_{,x}}+\left[\textcolor{black}{\varrho\left(\varrho+1 \right)}-\frac{ \kappa^{2} }{1-x^2} \right]\psi_{0}=0,  \label{eq9}
\end{eqnarray}
where
\begin{eqnarray*}
\varrho=\frac{1}{2}\sqrt{1-4\varepsilon\slashed{\rho}_{0}^{2}}-\frac{1}{2},\quad \kappa=\frac{s\slashed{\rho}_{0}}{b}.
\end{eqnarray*}
This is the associated Legendre differential equation. \textcolor{black}{Hence, its regular solution is given in terms of the associated Legendre function ($\mathcal{P}^{\kappa}_{\varrho}$), as $\psi_{0}\left(x \right)=\mathcal{N}\mathcal{P}^{\kappa}_{\varrho}\left(x \right)$ where $\mathcal{N}$ is a constant. \textcolor{black}{The solution function $\mathcal{P}^{\kappa}_{\varrho}\left(x \right)$ can become polynomial of degree $n$ with respect to $x$ provided $\varrho=n$. This condition results in the following expression
\textcolor{black}{\begin{eqnarray}
\mathcal{E}_{ns}=-s \hbar\omega_{rf}  \pm \hbar c \sqrt{\frac{m^2c^2}{\hbar ^{2}}-\frac{n\left( n+1\right) }{\slashed{\rho}_{0}^2}},    \label{eq10}
\end{eqnarray}}
for the spectrum of energy ($\mathcal{E}_{ns}$). Accordingly, by using the Eq. (\ref{f-comp}), one can obtain the spinor in the Eq. (\ref{eq1}) as follows}
\textcolor{black}{\begin{eqnarray}
\Psi=\textrm{e}^{-i\frac{\mathcal{E}_{ns}}{\hbar} t}\textrm{e}^{is\phi}\,\mathcal{N}\,\left(\begin{array}{c}
\frac{\sqrt{x^2-1}}{\slashed{\rho}_{0}(\tilde{\lambda}-\tilde{m})}\left[\partial_{x}\,\mathcal{P}^{\kappa}_{\varrho}\left(x \right)\right]-\frac{s/b}{\sqrt{x^2-1}}\mathcal{P}^{\kappa}_{\varrho}\left(x \right) \\
\mathcal{P}^{\kappa}_{\varrho}\left(x \right)\\
\mathcal{P}^{\kappa}_{\varrho}\left(x \right) \\
-\frac{\sqrt{x^2-1}}{\slashed{\rho}_{0}(\tilde{\lambda}+\tilde{m})}\left[\partial_{x}\,\mathcal{P}^{\kappa}_{\varrho}\left(x \right)\right]+\frac{s/b}{\sqrt{x^2-1}}\mathcal{P}^{\kappa}_{\varrho}\left(x \right)
\end{array}\right).\label{field-EWH}
\end{eqnarray}}
At first look, it can be seen that the energy spectrum for relativistic vector bosons in the rotating frame of the considered negative curvature wormholes are quite similar (or same if the radius of the curvature of the hyperbolic wormhole and elliptic wormhole are the same, $\rho_{0}=\slashed{\rho}_{0}$). The previous discussions in the section \ref{sec3.1} also apply in this scenario by considering $\rho_{0}\longrightarrow \slashed{\rho}_{0}$.}

\section{Summary and discussions}\label{sec4}

In this manuscript, we have studied the dynamics of relativistic \textcolor{black}{vector bosons} in the \textcolor{black}{rotating frame} of \textcolor{black}{negative curvature wormholes} by solving the corresponding fully-covariant \textcolor{black}{vector boson equation}. First of all, we have derived a general non-perturbative second-order wave equation for the considered systems, and we obtained exact solutions of the wave equation in two different scenarios. By assuming the \textcolor{black}{negative curvature wormholes are} \textcolor{black}{hyperbolic wormhole} and \textcolor{black}{elliptic wormhole}, we arrive at exact energy expressions, respectively. More interestingly, we have found that the obtained energy spectra in each case (see Eqs. (\ref{eq7}) and (\ref{eq10})) are  the same  provided the radius of the curvature of the considered \textcolor{black}{wormholes} are the same. Our results have shown that the energy of the system in questions depends explicitly on the radius of the curvature of the \textcolor{black}{wormholes}, and the angular frequency of the uniformly \textcolor{black}{rotating frame}. However, these energy spectra are independent from the radius of the \textcolor{black}{wormholes}, even though the solution functions depend on these parameters. We have observed that the coupling between the angular frequency of the uniformly \textcolor{black}{rotating frame} and the particle's spin is responsible for symmetry breaking around the zero energy \textcolor{black}{\cite{New}}. \textcolor{black}{Furthermore, it is evident that when the systems are in the ground state, the relativistic energy gives the value obtained by adding or substracting the particle's rest mass energy with the energy arising from the coupling between the angular frequency of the rotating frame and the particle's spin (see also \cite{exp1,exp2,exp3}).} Also, at first look, it is clear that our results can be adapted to massless \textcolor{black}{vector bosons} without loss of generality. In this case, our results imply that one can measure only the energy contribution originating from the mentioned spin-angular frequency coupling if the system is in the ground state. Moreover, one can see that the considered systems cannot be stable for each excited state if the \textcolor{black}{vector bosons} are massless ($m=0$). This is because the resulting energy expressions become complex-valued especially for excited states. \textcolor{black}{Consequently, these systems cannot remain stable, because the corresponding modes undergo decay or growth determined by the sign of the energy's imaginary part, leading to their evolution over time (note that $\Psi\propto \textrm{e}^{-i\,\frac{\mathcal{E}_{ns}}{\hbar} \, t}$ }. Here, by considering the $\tilde{\rho}$ is the radius of the curvature of the \textcolor{black}{considered negative curvature wormholes}, we can write the obtained general decay time expression for the damped modes as the following (see also \cite{guvendi2022vector})
\textcolor{black}{\begin{eqnarray*}
\tau_{n}\rightsquigarrow \frac{\tilde{\rho}}{c\sqrt{n(n+1)}},\label{dt}
\end{eqnarray*} }
provided $n\geqslant 1$ and $m^2\rightsquigarrow 0$. Here, it should be noted that the $\tilde{\rho}$ is in units of length. These results show that the decay time of the damped modes can be very long (short) if the $\tilde{\rho}$ is large (small). Also, it is clear that the duration of the decay process is dominated by the first excited state ($n=1$) because the others decay faster than the $n=1$ \textcolor{black}{state. Also, in principle, the results given by the Eqs. (\ref{eq7}), and (\ref{eq10}) imply that the relativistic energy ($\mathcal{E}_{ns}$) equals the energy arising from the coupling between the particle's spin and the rotating frame's angular frequency if the vector boson has a critical mass, $m_{c} = \frac{\hbar}{c\tilde{\rho}}\sqrt{n(n+1)}$. This is because the exact energy spectrum (unified by considering $\rho_{0}=\slashed{\rho}_{0}=\tilde{\rho}$) for the considered systems is in the following form
\begin{eqnarray*}
\mathcal{E}_{ns}=-s \hbar\omega_{rf}  \pm \hbar c \sqrt{\frac{m^2c^2}{\hbar ^{2}}-\frac{n\left( n+1\right) }{\tilde{\rho}^2}},
\end{eqnarray*}
and becomes $\mathcal{E}_{ns}\rightsquigarrow -s\,\hbar\,\omega_{rf}$ when $m\rightsquigarrow m_{c}$.}

\section*{\small{Acknowledgement}}
\textcolor{black}{The authors acknowledge the referees for their helpful suggestions and rigorous evaluation.}
\section*{\small{Funding}}
No fund was received for this research.
\section*{\small{Data Availability Statement}}
No Data associated in the manuscript.
\section*{\small{Conflict of Interest Statement}}
No conflict of interest has been declared by the authors.

\end{document}